\makeatletter
\@ifundefined{@parse@version@dash}{%
\def\@parse@version#1{\@parse@version@0#1}
\def\@parse@version@#1/#2/#3#4#5\@nil{%
\@parse@version@dash#1-#2-#3#4\@nil}
\def\@parse@version@dash#1-#2-#3#4#5\@nil{%
  \if\relax#2\relax\else#1\fi#2#3#4 }
}{} 
\makeatother
 
\documentclass[%
 reprint,
superscriptaddress, 
nofootinbib,
 amsmath,amssymb,    
 aps,
 ]{revtex4-2}

\usepackage{color}
\usepackage{graphicx}
\usepackage{dcolumn}
\usepackage{bm}
\RequirePackage{siunitx}   		
\DeclareSIUnit\eVperc{\eV\per\clight}
\DeclareSIUnit\clight{\text{\ensuremath{c}}}
\RequirePackage{booktabs}  
\RequirePackage{multirow}		
\RequirePackage{pgf}
\RequirePackage{hyperref}
\RequirePackage{physics}

\newcommand{\nuc}[2]{{}^{#1}{\textnormal{#2}}}

\begin{document} 


 \title{Fine-tunings in nucleosynthesis and the emergence of life: Status and perspectives}

\author{Ulf-G.~Mei{\ss}ner}
\email{meissner@hiskp.uni-bonn.de}%
\affiliation{Helmholtz-Institut~f\"{u}r~Strahlen-~und~Kernphysik,%
~Rheinische~Friedrich-Wilhelms~Universit\"{a}t~Bonn,~D-53115~Bonn,~Germany}
\affiliation{Bethe~Center~for~Theoretical~Physics,%
~Rheinische~Friedrich-Wilhelms~Universit\"{a}t~Bonn,~D-53115~Bonn,~Germany}
\affiliation{Center~for~Science~and~Thought,%
~Rheinische~Friedrich-Wilhelms~Universit\"{a}t~Bonn,~D-53115~Bonn,~Germany}
\affiliation{Institute~for~Advanced~Simulation~(IAS-4),%
~Forschungszentrum~J\"{u}lich,~D-52425~J\"{u}lich,~Germany}
\affiliation{Peng Huanwu Collaborative Center for Research and Education,
International Institute for Interdisciplinary and Frontiers, Beihang University, Beijing 100191,
China}

%
\author{Bernard~Ch.~Metsch}%
\email{metsch@hiskp.uni-bonn.de}%
\affiliation{Institute~for~Advanced~Simulation~(IAS-4),%
~Forschungszentrum~J\"{u}lich,~D-52425~J\"{u}lich,~Germany}
\affiliation{Helmholtz-Institut~f\"{u}r~Strahlen-~und~Kernphysik,%
~Rheinische~Friedrich-Wilhelms~Universit\"{a}t~Bonn,~D-53115~Bonn,~Germany}

\author{Helen~Meyer}%
\email{hmeyer@hiskp.uni-bonn.de}%
\affiliation{Helmholtz-Institut~f\"{u}r~Strahlen-~und~Kernphysik,%
~Rheinische~Friedrich-Wilhelms~Universit\"{a}t~Bonn,~D-53115~Bonn,~Germany}
\affiliation{Bethe~Center~for~Theoretical~Physics,%
~Rheinische~Friedrich-Wilhelms~Universit\"{a}t~Bonn,~D-53115~Bonn,~Germany}

\date{\today}

\begin{abstract}
We discuss the fine-tunings of nuclear reactions in the Big Bang
and in stars and draw some conclusions on the emergence of the
light elements and the life-relevant elements carbon and oxygen. 
We also stress how to improve these calculations in the future. This
requires a concerted effort of different communities, especially in 
nuclear reaction theory, lattice QCD for few-nucleon systems, 
stellar evolution calculations, particle physics and philosophy.
\end{abstract}

\maketitle


\section{Prologue}
\label{sec:pro}

This viewpoint grew out of discussions with Christian Caron concerning
the Advanced Grant from the European Research Council on ``Emergent complexity
from strong interactions'', in particular the third work package on
``How fine-tuned is nucleosynthesis?''. This is a topic that can only
be addressed in theory, as Nature gives us specific values for the
fundamental constants that govern the emergence of nucleons and nuclei,
and thus the emergence of life as we know it.
The aim of this viewpoint is to get  more people interested in such type
of research, which also has some overlap with philosophy --
for recent discussions see~\cite{DeVuyst:2020wzv,Goff:2024wvq}.

\section{Introduction}
\label{sec:intro}

The matter we are made of consists almost completely of atomic nuclei.
These are generated shortly after the Big Bang and in stars -- that is why one often
says that we are made of stardust. Of particular relevance are the
$^{12}$C and $^{16}$O nuclei, which form the basis of the life on
Earth as we know it. These elements are generated in hot, old stars
through the triple-alpha process and the radiative capture process
$^{12}{\rm C}(\alpha,\gamma)^{16}{\rm O}$, respectively. Both $^{12}$C
and $^{16}$O are alpha-type nuclei, that is to a good approximation they
can be described as bound states of three, respectively four, $^4$He
particles. $^4$He is already generated abundantly shortly after the Big Bang,
and it is well-known that the triple-alpha reaction features the
Hoyle state~\cite{Hoyle:1954zz}, a resonance close to the $^4$He+$^8$Be threshold,
that is required to form a sufficient  amount of carbon and oxygen in stars
to enable life on Earth. Similarly, in Big Bang nucleosynthesis (BBN), the
deuterium bottleneck plays a crucial role, as the Universe has to cool 
down sufficiently so that the neutron-proton fusion to deuterium is not undone
by the abundant energetic photons (with $E_\gamma > 2.2$~MeV) that disintegrate   
the deuteron.\footnote
{This, by the way, is very different from high-energy heavy ion collisions
on Earth, where  photons play essentially no role. The photons in the Big Bang
are due to particle-antiparticle annihilations happening before element synthesis.}
Therefore, to understand
these fine-tunings and draw conclusions on possible variations of the
fundamental constants is not only interesting by itself, it also is
a necessary requirement for the study of possible Beyond the Standard
Model (BSM) effects in these reactions.  The fundamental parameters under
consideration are related to the various interactions pertinent to nuclear 
physics, namely the light quark masses $m_u,m_d,m_s$ related to the strong interactions, 
the electromagnetic fine-structure constant $\alpha_{\rm EM}$, the coupling constant
of QED,  and, to a  lesser extent, the Fermi
coupling constant $G_F$ that gives the strength of the weak interactions at
low energies. For a study of a Universe without weak interactions, see Ref.~\cite{Harnik:2006vj}.
It should be stressed that the formation of the life-enabling elements
is a necessary but not sufficient condition for the emergence of life as we know
it -- it requires many other sciences to come to a complete picture, see e.g.~\cite{enzy}.
Having said that, 
we here concentrate on the nuclear physics aspects of this topic.
This paper is not a review, but we
rather intend to stress some recent developments and  loopholes in such type of calculations as well
as discussing required improvements in this intricate interplay of effective
field theories (EFTs), lattice QCD (LQCD) calculations and nuclear reaction modeling.

The paper is organized as follows: In Sect.~\ref{sec:BBN} we
discuss the constraints on the electromagnetic fine-structure constant $\alpha_{\rm EM}$
and the light quark masses in the Big Bang.
Sect.~\ref{sec:stars} considers similar variations for various
nuclear reactions pertinent to the generation of carbon and oxygen.
A discussion and outlook is given in Sect.~\ref{sec:summary}.

\section{Fine-tunings in the Big Bang}
\label{sec:BBN}

The light elements up to $A=7$ are generated in BBN through an
intricate interplay of nuclear reactions, the so-called reaction
network, that is given in terms of coupled differential equations
(the rate equations)
for the abundances $Y_i = n_i/n_B$, with $n_i$ the density of nucleus $i$
and $n_B$ the total baryon density.

The first point we want to stress
is that one should use the different publicly available codes for BBN
to get an estimate of the systematical errors in the network calculation.
These codes are: 
\texttt{NUC123}~\cite{Kawano:1992ua},
\texttt{AlterBBN}~\cite{Arbey:2011nf,Arbey:2018zfh},
\texttt{PArthENoPE}~\cite{Pisanti:2007hk,Gariazzo:2021iiu},
\texttt{PRIMAT}~\cite{Pitrou:2018cgg} and
\texttt{PRyMordial}~\cite{Burns:2023sgx,PRyMordial-Code}.
These differ mainly in the number of reactions considered, in the
parametrization of the nuclear reactions and the numerical treatment
of the rate equations.

\begin{figure}[htb!]
\includegraphics[width=0.49\textwidth]{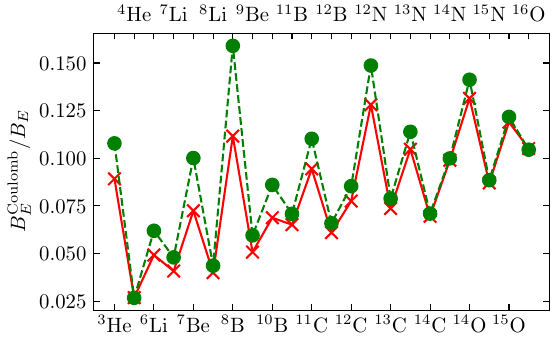}
\caption{Coulomb contribution to the nuclear binding energies based on NLEFT
(red crosses) compared to the Bethe-Weizs\"acker formula (green circles).}
\label{fig:EC}
\end{figure}

Let us first consider the variation of $\alpha_{\rm EM}$. There are
essentially four different ways a dependence on $\alpha_{\rm EM}$
is generated: (i) in the nuclear reactions rates, we encounter the
Coulomb barrier, which leads to an energy-dependent penetration
factor in the cross section~\cite{Blatt:1952ije},
(ii) radiative capture reactions, (iii) in the $n\leftrightarrow p$ conversion
and in $\beta$-decay rates, one has to deal with final and/or initial state
interactions between charged particles, and (iv) there are various
indirect effects generated by the Coulomb contribution to the nuclear
binding energies and the QED contribution to the neutron-proton mass
difference, $\Delta Q_n = Q_n^{\rm QED} \cdot \delta\alpha_{\rm EM} =
-0.58(16)~{\rm MeV}\cdot \delta \alpha_{\rm EM}$~\cite{Gasser:2020mzy}.
An up-to-date calculation of the Coulomb contribution to the
nuclear binding energies based on Nuclear Lattice Effective Field
Theory (NLEFT)~\cite{Elhatisari:2022zrb} compared to the time-honored
Bethe-Weizs\"acker formula \cite{Weizsacker:1935bkz,Bethe:1936zz} is
displayed in Fig.~\ref{fig:EC}, showing that more and more of the quantities
under consideration can be calculated {\em ab initio}.

The biggest source of uncertainty are indeed the reaction rates and
cross sections, with the exception of the $n+p\to d+\gamma$ reaction, that
can be precisely calculated in pionless EFT~\cite{Rupak:1999rk}.
We remark in passing that at temperatures (or energies) relevant in primordial nucleosynthesis
the relevant reaction rates can be measured. In Ref.~\cite{Meissner:2023voo}, we made use of available
experimental data to find novel parametrizations of the leading 17 reaction rates.
Together with the pionless EFT rate for $n+p\to d+\gamma$, we implemented these new rates into the
reaction network of the 5 codes mentioned above to draw conclusions on the allowed variation of
$\alpha_{\rm EM}$ from the reliable measurements of the $d$ and $^4$He abundances (the $^7$Li abundance
features the so far unsolved Lithium puzzle~\cite{Fields:2011zzb}).
First, we observed that the different codes gave rather similar results. Second,
we found  $|\delta\alpha_{\rm EM}| <1.8\%$ from $^4$He and  $|\delta\alpha_{\rm EM}| <0.8\%$
from $d$, which are tighter constraints than found previously.

To partly overcome the modeling of the nuclear reactions, in Ref.~\cite{Meissner:2024pgh},
we used Halo-EFT (for a review, see~\cite{Hammer:2017tjm}) to describe the reactions
$n + {}^7{\rm Li} \to  {}^8{\rm Li} + \gamma$~\cite{Fernando:2011ts,Higa:2020kfs},
$p + {}^7{\rm Be} \to  {}^8{\rm B} + \gamma$~\cite{Higa:2022mlt,Zhang:2017yqc},
${}^3{\rm H} + {}^4{\rm He} \to  {}^7{\rm Li} + \gamma$,
${}^3{\rm He} + {}^4{\rm He} \to  {}^7{\rm Be} + \gamma$~\cite{Higa:2016igc,Premarathna:2019tup,Zhang:2019odg}. Using these rates, one finds  substantial deviations from the
$\alpha_{\rm EM}$-dependence of the parameterized rates obtained for these
reactions previously, however, the impact on the resulting abundances and on
their $\alpha_{\rm EM}$-dependence of the light elements $\nuc{2}{H}$,
$\nuc{3}{H}+\nuc{3}{He}$, $\nuc{4}{He}$, $\nuc{6}{Li}$ is very minor only.  In
contrast for the $\nuc{7}{Li}+\nuc{7}{Be}$-abundance we do find that
the $\alpha_{\rm EM}$-dependence differs appreciably from that of the previous
parameterized results, this $\alpha_{\rm EM}$-dependence being much more
pronounced and clearly non-linear with the Halo-EFT rates.  Also the
nominal abundance (i.e. calculated with the current value of
the fine-structure constant) of $\nuc{7}{Li}+\nuc{7}{Be}$ is
larger by almost 10 \%, whereas the other abundances remain
practically unchanged. For reactions involving charged particles,
the Halo-EFT calculation accounts for the charged particle repulsion
by inclusion of the full Coulomb propagator in all reaction steps.
As shown in~\cite{Meissner:2024pgh} these Coulomb effects cannot always be
approximated by a universal penetration factor. It was also found that in some
cases the study of the $\alpha_{\rm EM}$ dependence of cross sections and the
corresponding rates within the framework of Halo-EFT is limited by singularities
appearing in the normalization, that enters as a factor in the
resulting cross sections. This was found to be relevant for the
${}^{3}{\rm He}+{}^{4}{\rm He} \to {}^{7}{\rm Be}+\gamma$ reaction,
limiting the study to relative variations of $\alpha_{\rm EM}$ to
less than $6\%$\,. We will come back  to the microscopic description of
the nuclear reactions in the following sections.

\begin{figure}[htb!]
\includegraphics[width=0.47\textwidth]{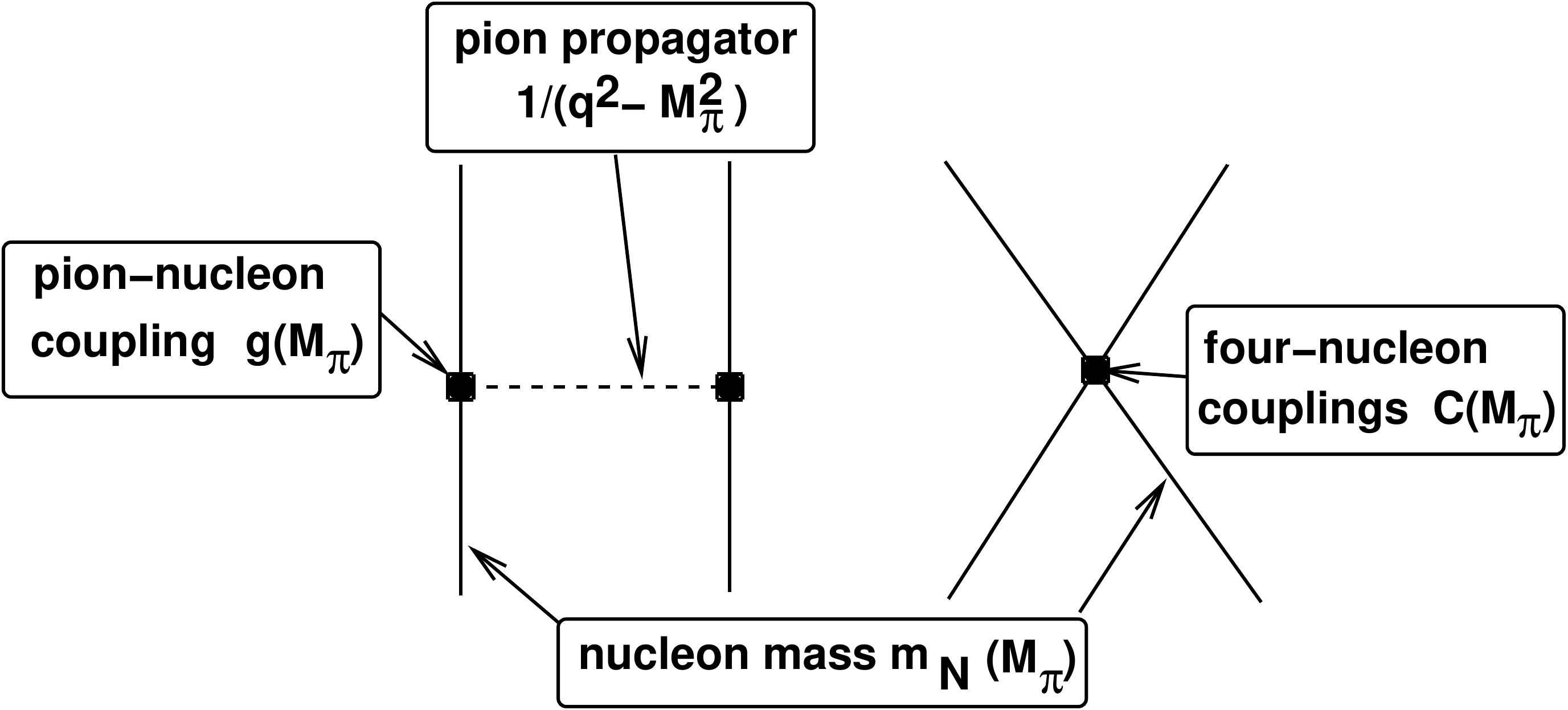}
\caption{Pion mass dependence of the leading order nucleon-nucleon interaction.
Left: one-pion exchange. Right: Leading four-nucleon contact interactions. Solid (dashed)
lines denote nucleons (pions).}
\label{fig:Mpi}
\end{figure}

Next we consider the variations of the light quark masses $m_u,m_d$, which 
have been considered in many works using various levels of modeling. The
first work that used pionless EFT to address the issue was 
Ref.~\cite{Bedaque:2010hr}, followed by the use of chiral nuclear EFT with some
resonance saturation modeling of the four-nucleon contact terms in Ref.~\cite{Berengut:2013nh}.
Using the Gell-Mann--Oakes--Renner relation, the quark mass dependence can be
mapped onto the pion mass dependence and the leading nuclear interactions depend
explicitly and implicitly on the pion mass $M_\pi$ as shown in Fig.~\ref{fig:Mpi}.
These works came to the conclusion that variations of the light quark mass of about 1\% are
consistent with the observed abundances, where the $^4$He abundance sets a tighter
constraint than the $d$ abundance (and similarly for variations of the Higgs VEV $v$. Remember
that for constant Yukawa couplings, variations in the quark mass and $v$ are the same).
Fig.~\ref{fig:Mpi} already exhibits all the ingredients and complications that such type of
calculation exhibits. On the one hand, LQCD easily provides data for the nucleon mass 
and the pion-nucleon coupling constant $g$ (using  the Goldberger-Treiman relation) at
varying quark (pion) masses,
whereas the four-nucleon couplings $C_{S,T}$ can not so easily be obtained from
LQCD. This is where the low-energy theorems (LETs) of Refs.~\cite{Baru:2015ira,Baru:2016evv}
enter, as they allow to connect LQCD data for the two-nucleon system at relatively
large pion masses with the small changes of the pion mass around its physical value
as considered e.g. in BBN. Variations of the Higgs VEV $v$ came back into focus
recently in Ref.~\cite{Burns:2024ods} due to the newly published EMPRESS data~\cite{Matsumoto:2022tlr}
for the $^4$He abundance, using one-boson-exchange modeling. Here, the $^4$He
abundance led to the strongest constraints, somewhat stronger than found earlier.
The variation of $\alpha_{\rm EM}$ in view of the EMPRESS data was considered in
Ref.~\cite{Seto:2023yal}.
The bounds on $\delta v/v$ were reconsidered in Ref.~\cite{Meyer:2024auq}, where LQCD data for $m_N (M_\pi)$ and
$g_A(M_\pi)$ were used together with LQCD data on the deuteron binding energy and the
S-wave scattering lengths $a_s,a_t$ for pion masses from 300 to 800~MeV combined with the
above-mentioned LETs. The most important finding in that paper was that the backwards
reaction $d + \gamma \to n+p$ is influenced by a change in the deuteron binding energy.
Because of the deuterium bottleneck, the rate of this reaction defines the beginning of BBN
and has therefore a sizeable impact especially on the $^4$He abundance. The reaction rate for
$d + \gamma \to n+p$ is derived from the rate for $n+p\to d+\gamma$ through a detailed
balance relation
\begin{eqnarray}
  &&  \expval{\sigma(d+\gamma\to n+p)v} \nonumber\\
  && = \aleph \, T_9^{3/2} \exp(\kappa/T_9)\,
    \expval{\sigma(n+p\to d + \gamma)v}~,
\end{eqnarray}
where $T_9$ is the photon temperature in units of $10^9\si{\K}$. The parameter
$\aleph$ depends on the deuteron mass (and hence on the deuteron binding energy) and the
neutron and proton mass through
$\aleph \propto ({m_n m_p}/{m_d})^{3/2}$, 
while the parameter $\kappa$ corresponds to the reaction $Q$-value and is, for this reaction,
directly proportional to the deuteron binding energy $\kappa \propto B_d$. When making changes
to the deuteron binding energy (and hence its mass) and the nucleon mass, this needs to be taken into
account and it has a significant effect on the $d$ and $^4$He abundances. Even more, the deuterium
abundance now sets the strongest constraints, comparing with the PDG numbers leads to the
$2\sigma$ bound:
\begin{equation}\label{eq:v}
-0.07\% \leq \delta v/v \leq -0.02\%~,
\end{equation}
which is a much stronger fine-tuning than observed in all other earlier works. The corresponding
$^4$He bound is $-0.7\% \leq \delta v/v \leq +0.4\%$.

Recently, the dependence on the strange quark mass $m_s$ was also studied. This is more challenging,
as strange quark effects are mostly indirect and chiral EFTs including baryons and kaons  encounter
convergence problems for a number of observables. In Ref.~\cite{Meissner:2025jfs} it was argued that
the dominant strangeness effect in BBN is the strange quark contribution to the nucleon mass shift,
parameterized in terms of the strangeness $\sigma$-term, $\sigma_s =\langle N| m_s\bar{s}s|N\rangle$.
Varying the nucleon mass in the leading eight reactions that involve neutrons, protons and the four
lightest nuclei, using the various BBN network codes, leads to strict limits on the allowed nucleon
mass variations that translate into an upper bound on possible variations of the strange quark mass,
$|\Delta m_s/m_s| \leq 5.1\%$ (assuming the strange quark condensate not to vary).
It is amusing to note that a mere $2\%$ reduction of the nucleon mass
would solve the Lithium problem. For a general discussion of the primordial nuclear abundances on
fundamental nuclear observables such as binding energies, scattering lengths, neutron lifetime, etc.,
see~\cite{Meissner:2022dhg}.

\section{Fine-tunings in stellar nucleosynthesis}
\label{sec:stars}

Now we consider the generation of carbon and oxygen in hot, old stars. Here, the triple-$\alpha$
process exhibits two fine-tunings, namely the closeness of the instable but long-lived $^8$Be
nucleus to the $2\alpha$ threshold and of the Hoyle state to the $3\alpha$ threshold. It was speculated
already by Weinberg (and others)  that these two fine-tunings are correlated~\cite{SWface}
(also, such correlations are implicit in the ground-breaking work of Ikeda and collaborators on
alpha-clustering~\cite{Ikeda}).
That this is indeed the case
could only be shown  using NLEFT in Refs.~\cite{Epelbaum:2012iu,Epelbaum:2013wla} a decade later.

Let us concentrate on the closeness of the Hoyle state to the $3\alpha$ threshold. It is well-known
that the rate of the triple-alpha process depends exponentially on the energy
difference $E_R =E_{12}^*-3E_4^{}$,
where $E_{12}^*$ is the excitation energy of the Hoyle state and $E_4$ the mass of the alpha-particle.
First numerical experiments by varying  this energy difference in stellar simulations were
performed in Refs.~\cite{Livio:1989,Oberhummer:2000zj}. This was improved in~\cite{Huang:2018kok},
were a larger range of star masses, $M_\star = (15-40)M_\odot$ and also  both solar and low metallicity
were considered. Furthermore, these authors also investigated the generation of carbon, oxygen and
heavier elements. They find values for $\Delta E_R$ that depend on the metallicity.
For low metallicity, for negative values of $\Delta E_R$, carbon production
limits this to $\Delta E_R \geq -300\,$keV and for positive values, oxygen production leads to
$\Delta E_R \leq 300\,$keV. For solar metallicity, these bounds are found to be narrower by
a factor of 2, but these are still larger than found in the earlier
studies ~\cite{Livio:1989,Oberhummer:2000zj}. The updated NLEFT analysis~\cite{Lahde:2019yvr}, which for the
first time used the LETs of Refs.~\cite{Baru:2015ira,Baru:2016evv} in the pion mass dependence
of the four-nucleon operators, gave a range of possible quark mass variations related to
the treatment of the short-distance two-nucleon physics, leading to bounds between $0.4\%$
to $5\%$ for $\delta m_q/m_q$ (or $\delta v/v$). The corresponding bounds on the variation
of $\alpha_{\rm EM}$, $|\delta\alpha_{\rm EM}/\alpha_{\rm EM}| < 7.5\%$, are also less strict
than found in BBN. Interestingly, the nuclear reaction rates in the Big Bang and in stars
limit the mysterious QCD $\theta$-parameter to $\theta \lesssim 0.1$~\cite{Lee:2020tmi},
which is not particular fine-tuned and far away from the experimental limit of $\theta \simeq 10^{-11}$.

These type of calculations could be improved
in two aspects. First, one would like to not only vary $E_R$ but all the masses, reactions
rates etc. in the stellar burning processes, similar to what is done in BBN. That, however, is out of reach
of present computational capabilities. Second, more LQCD data for the two-nucleon system at lower
pion masses could certainly help to reduce the uncertainty of the allowed variations in the quark masses
(the Higgs VEV) derived from the variations found for $E_R$ in the stellar modeling.

Another venue to address the issue of parameter variations in the triple-alpha process
has recently become available in the framework of NLEFT. After the pioneering work
on $\alpha$-$\alpha$ scattering at N2LO published in 2015~\cite{Elhatisari:2015iga}, it was possible
to study the dependence on the fundamental constants of the SM of this first reaction
in the triple-alpha process~\cite{Elhatisari:2021eyg}. It was found that positive shifts in
the pion mass have a small effect on the S-wave phase shift, whereas lowering the pion mass
adds some repulsion in the two-alpha system. The effect on the D-wave phase shift turns out to
be more pronounced as signaled by the D-wave resonance parameters. Variations of $\alpha_{\rm EM}$
have almost no effect on the low-energy $\alpha$-$\alpha$ phase shifts.  This calculation
can clearly be improved by going to N3LO using the high-fidelity chiral forces
from~\cite{Elhatisari:2022zrb}. In a next step, one would extend such type of study to the
second reaction in the triple-alpha process, namely the fusion of $^4$He with $^8$Be to generate
$^{12}$C via the Hoyle state. This would be followed by a study of the holy grail of
nuclear astrophysics, $^{12}{\rm C}(\alpha,\gamma)^{16}{\rm O}$, using the same methods.
First unpublished results on elastic $\alpha$-${}^{12}$C scattering in S- and P-wave look indeed
promising, paving the way for the calculation of the radiative capture reaction.
It would also be interesting to use the pionless (cluster) EFT of
Refs.~\cite{Ando:2022flx,Ando:2023lrt,Ando:2025ibj} to study the dependence of
$\alpha$-${}^{12}$C on $\alpha_{\rm EM}$ and the light quark masses.

Other fine-tunings observed in element generation or energy production in stars
are  the proton capture in $^{56}$Ni to produce heavier elements in X-ray
bursts~\cite{Schatz:1998zz}, the $^{205}$Pb - $^{205}$Tl conversion, as  $^{205}$Pb
plays a major role in revealing the formation history of the sun~\cite{Leckenby:2024pmj}  and
electron capture in $^{20}$Ne, that has a decisive impact on the evolution of the core
for stars with $7$-$11$ solar masses~\cite{Kirsebom:2019tjd}. To our knowledge, these
have not been scrutinized along the lines discussed before.

\section{Discussion and outlook}
\label{sec:summary}

Now we are at the point to draw some conclusions on the fine-tuning (for a general discussion
relating it to the anthropic principle, see~\cite{Meissner:2014pma}, or the multiverse, see
\cite{Donoghue:2016tjk}). As can be seen, the bounds
on possible variations of the Higgs VEV are much stronger from BBN, and also the theoretical 
uncertainty on the variations deduced from carbon and oxygen production are larger. In particular,
the sub-percent variations for $v$, cf. Eq.~\eqref{eq:v}, sheds new light on the fine-tunings in BBN,
and it also serves as a benchmark for  BSM models applied to primordial nucleosynthesis.

There are a number of issues that need to be tackled to make these calculations more conclusive
and/or  can lead to novel insights:
\begin{itemize}
\item It would be very valuable to use more reaction theory calculations in BBN to study their parameter
dependence. In case of the largely analytical Halo-EFT, this can be done by other researchers than the authors
of the various papers, as discussed before. This is very
different from the no-core-shell model coupled to the continuum, that offers a different {\em ab initio}
approach the nuclear reactions in the BBN network to what has been discussed before, see e.g.
Refs.~\cite{Hupin:2018biv,Kravvaris:2022eyf,Hebborn:2022iiz,Atkinson:2024zrm}.  The proponents
of this approach should consider performing the pertinent calculations. Of course, NLEFT will
also contribute to these developments.
\item Cluster models based on EFT approaches can also be used to get a handle on possible
parameter variations. In particular, the recent work of Refs.~\cite{Ando:2022flx,Ando:2023lrt,Ando:2025ibj}
on the holy grail of nuclear astrophysics should be mentioned, where the $\alpha_{\rm EM}$ dependence is 
explicit and the one on the quark masses implicit, in terms of the pertinent scattering parameters and nuclear masses.
\item As already stated, it would be very valuable to have LQCD simulations for the two-nucleon system
at lower quark masses, which would help to reduce the uncertainty generated from the quark mass
expansion of the four-nucleon contact terms, that so far is least constrained.
\item Finally, we also mention that so far most calculations keep the Yukawa couplings constant.
It would be very interesting to perform  more work along the lines of Ref.~\cite{Coc:2006sx}, where
the interrelations between the fundamental parameters arising in unified theories were considered.
\end{itemize}
We hope that with the discussion presented here, more researchers from the different fields
mentioned will be contributing to this highly interesting topic. Finally, it would also be important to
bring the somewhat disjoint communities of philosophers and physicists closer together to deepen
our understanding of this topic.

\begin{acknowledgments}

  We are grateful to Karlheinz Langanke for some useful remarks. 
  This project is part of the ERC Advanced Grant ``EXOTIC'' supported
  the European Research Council (ERC) under the European Union's Horizon
  2020 research and innovation programme (grant agreement
  No. 101018170), We further acknowledge support by
  the Chinese Academy of Sciences (CAS) President's International
  Fellowship Initiative (PIFI) (Grant No. 2025PD0022).
\end{acknowledgments}

\end{document}